\documentclass[aps,nofootinbib,prd,class-pre]{revtex4} 
\usepackage{amsmath,amssymb,amsfonts,amsthm}
\usepackage{graphicx}

\usepackage{subeqnarray}

\usepackage{subfigure}

\renewcommand{\v}[1]{\ensuremath{\mathbf{#1}}} 
\newcommand{\gv}[1]{\ensuremath{\mbox{\boldmath$ #1 $}}}

\newcommand{\ket}[1]{\left| #1 \right>} 

\newcommand{\be}{\begin{equation}}
\newcommand{\ee}{\end{equation}}
\newcommand{\bea}{\begin{eqnarray}}
\newcommand{\eea}{\end{eqnarray}}

\def\lab{\label}

\def\pa{\partial}

\def\rar{\rightarrow}

\def\al{\alpha}

\def\ga{\gamma}

\def\de{\delta}

\def\ep{\epsilon}

\def\la{\lambda}

\def\om{\omega}
\def\Om{\Omega}
\begin{document}

\title{Water Bridging Dynamics of Polymerase Chain Reaction\\ in the Gauge Theory Paradigm of Quantum Fields\footnote{{\em Water} {\bf 2017}, {\em 9}, 339. DOI:10.3390/w9050339}}


\author{L. Montagnier$^{1}$, J. A\"issa$^{2}$, A. Capolupo$^{3}$, T. J. A. Craddock$^{4}$, P. Kurian$^{5}$,\\ C. Lavallee$^{1}$, A. Polcari$^{6}$, P. Romano$^{6}$, A. Tedeschi$^{7}$,
and G. Vitiello$^{3}$}   

\address{%
$^{1}$  World Foundation for AIDS Research and Prevention, Paris; 1002175007mm@gmail.com \\
$^{2}$  Nanectis Biotechnologies, Jouy-en-Josas, France; 1002175007mm@gmail.com\\
$^{3}$  Dipartimento di Fisica "E.R. Caianiello", Universit\'a di Salerno and INFN, 84084 Fisciano (Salerno), Italy; capolupo@sa.infn.it ~~ vitiello@sa.infn.it\\
$^{4}$  Departments of Psychology and Neuroscience, Computer Science, and Clinical Immunology, and Clinical Systems Biology Group, Institute for Neuro-Immune Medicine, Nova Southeastern University, Fort Lauderdale, FL 33314, USA; tcraddock@nova.edu\\
$^{5}$  National Human Genome Center and Department of Medicine, Howard University College of Medicine, Washington, DC 20059, USA; Computational Physics Laboratory, Howard University, Washington, DC 20059, USA; pkurian@howard.edu\\
$^{6}$  Dipartimento di Scienze e Tecnologie,
Universit\'a del Sannio, Benevento, Italy and SPIN-CNR, Universit\'a  di Salerno, Fisciano (SA), Italy; a.polcari@libero.it ~~ promano@unisannio.it\\
$^{7}$  WHITE Holographic Bioresonance, Milano, Italy; gowhite@usa.net\\
}




\maketitle


{\bf Abstract} We discuss the role of water bridging the DNA-enzyme interaction by resorting to recent results showing that London dispersion forces between delocalized electrons of base pairs of DNA are responsible for the formation of dipole modes that can be recognized by \textit{Taq} polymerase.
We describe the dynamic origin of the high efficiency and precise targeting of \textit{Taq} activity in PCR.
The spatiotemporal distribution of interaction couplings, frequencies, amplitudes, and phase modulations comprise a pattern of fields which constitutes   the electromagnetic image of DNA in the surrounding water, which is what the polymerase enzyme actually recognizes in the DNA water environment.
The experimental realization of PCR amplification, achieved through replacement of the DNA template by the treatment of pure water with electromagnetic signals recorded from viral and bacterial DNA solutions, is found consistent with the gauge theory paradigm of quantum fields.

\vspace{0.7cm}








\section{Introduction}

In this paper, by resorting to recent results obtained in Refs.~\cite{watermediated} and ~\cite{Kurian},  we study the fundamental role of water dipole dynamics bridging the interaction between DNA and enzymatic proteins so as to allow polymerase chain reaction (PCR) processes to occur.
Some of us have previously reported~\cite{DNA1a,DNA1,DNA2a,DNA2}  the {\it in vitro} transduction of bacterial DNA sequences in  water  and in tumor  cells.
In water, the DNA was detected by  PCR using the \textit{Taq} polymerase from the heat-tolerant strain 
\textit{T. aquaticus}.
This phenomenon raises several important questions from the theoretical  point of view: how is it that a bacterial DNA polymerase ({\it Taq}) can ``read'' water nanostructures imprinted by bacterial DNA sequences?
If so, is there a more general mechanism operating in living cells?
The present article offers an explanation by using the formalism of the gauge theory paradigm of quantum fields along the research lines developed in Refs.~\cite{watermediated, DNA2a,DNA2,PRA2006,PRL1988,NuclPhys85,NuclPhys86,Alfinito:2001mm}.

In Ref.~\cite{Kurian}, London dispersion forces between DNA base pairs and the dynamics of delocalized electrons are shown to be responsible for the formation of molecular dipole structure.
Fine tuning of collective dynamic modes and synchronization of dipole vibrational modes between spatially separated nucleotides and enzymatic molecular subunits  have been theorized to promote and sustain the catalytic activity of classes of enzymes which interact with DNA. Long-range molecular correlations in the chemistry of DNA-enzyme interactions, discussed in Ref.~\cite{watermediated}, appear to play a key role in understanding how to reconcile the high efficiency and precise targeting of enzymatic catalytic activity with the intrinsically stochastic kinematics of a large number of molecular components~\cite{Schuss,Cosic,HU}. The highly ordered pattern in the spatial molecular arrangements and the time ordered sequence of steps characterizing the catalytic activity indicate that biochemical methods and the traditional study of the dipole and multipole dynamic structure of the electronic quantum conformations, out of which the chemical properties of interacting molecules emerge, need to be supplemented by results in quantum field theory (QFT). These indeed account for the collective molecular long-range correlations, which are responsible for the highly synchronized, fine-tuned DNA-enzyme interaction.

Thus, in the framework of spontaneously broken gauge symmetry theories, which has proven itself to be the most reliable theoretical scheme able to predict the wide array of experimental observations ranging from condensed matter to elementary particle physics, it has been derived in~\cite{watermediated} a generalized model of the radiative dipole wave field mediating the molecular interactions between DNA and enzyme in water and governing the dynamics of {\it Taq} DNA polymerase in PCR processes.

The model is based on the QFT interaction paradigm, which describes the interaction between two systems by the mutual exchange of a mediating wave field or quantum, in analogy with the photon exchanged by two electric charges in quantum electrodynamics~\cite{Itz,Blasone2011} (Figure 1). The basic constraint in the model construction comes of course from the physical characterization of the system under study. In our case, we have the macromolecules of DNA and enzyme and the water molecules, whose number under standard biological conditions constitutes the supermajority of the system molecules. DNA, enzyme, and water molecules are endowed, as their distinctive physical property in this model, with configurations of electric dipoles. Our modeling is thus focused in a natural way on the radiative dipole interaction.

The wave field mediating DNA-enzyme interactions, of which \textit{Taq} amplifying DNA sequence in PCR represents one specific case, are proposed to be governed by the dipole wave field propagating through the water matrix in which DNA and enzymes are embedded (Figure 1). The mathematical formalism and the results of such a study are reported in~\cite{watermediated}. In the present paper our aim is to discuss developments in the applications to DNA-enzyme interactions with particular attention paid to the {\it Taq} DNA polymerase in PCR processes, providing an understanding based on dynamical field-theoretic grounds, thus supplementing more standard presentations of PCR modeling based on kinetic and stoichiometric methods~\cite{Schuss,Cosic,HU}.

The result of our analysis is that in PCR amplification processes, and more generally in DNA-enzyme interactions, the spatial and temporal distributions of charges~\cite{watermediated,Kurian,Chen}, interaction couplings, frequencies, amplitudes, and phase modulations~\cite{watermediated,Kurian} form a pattern of fields, that constitutes the electromagnetic (em) image of the DNA, in such a way that what the enzyme ``sees'', at the level of molecular biology, is such an em image of DNA in the surrounding water. The DNA and the enzyme ``see'' each other's em images by exchanging quanta of the radiative dipole waves induced by their presence in the water molecular matrix, which thus acts effectively as a {\it bridge} between the two (of course, until they are sufficiently close for water exclusion and direct binding to occur).

The structure of the paper is as follows. The {\it Taq}-DNA interaction in PCR processes, and DNA transduction in water are discussed in Section 2. The formalism developed in~\cite{watermediated} is summarized in Section 3. The description of PCR processes in terms of the gauge theory paradigm of quantum fields is presented in Section 4. Section 5 is devoted to further discussions and conclusions. In the Note added in proof, we comment on the first observation of a chiral spine of hydration templated by a biomolecule, obtained by use of chiral sum frequency generation spectroscopy \cite{McDermott},
in relation with the PCR analysis in terms of mediating water dynamics.
Details of the protocol for DNA transduction in water are reported in  Appendix A; some data on the dipole molecular structure of DNA and {\it Taq} in Appendix B and some mathematical details on the water dipole field in Appendix C.

\begin{figure*}
		 \includegraphics[width={0.3\textwidth}]{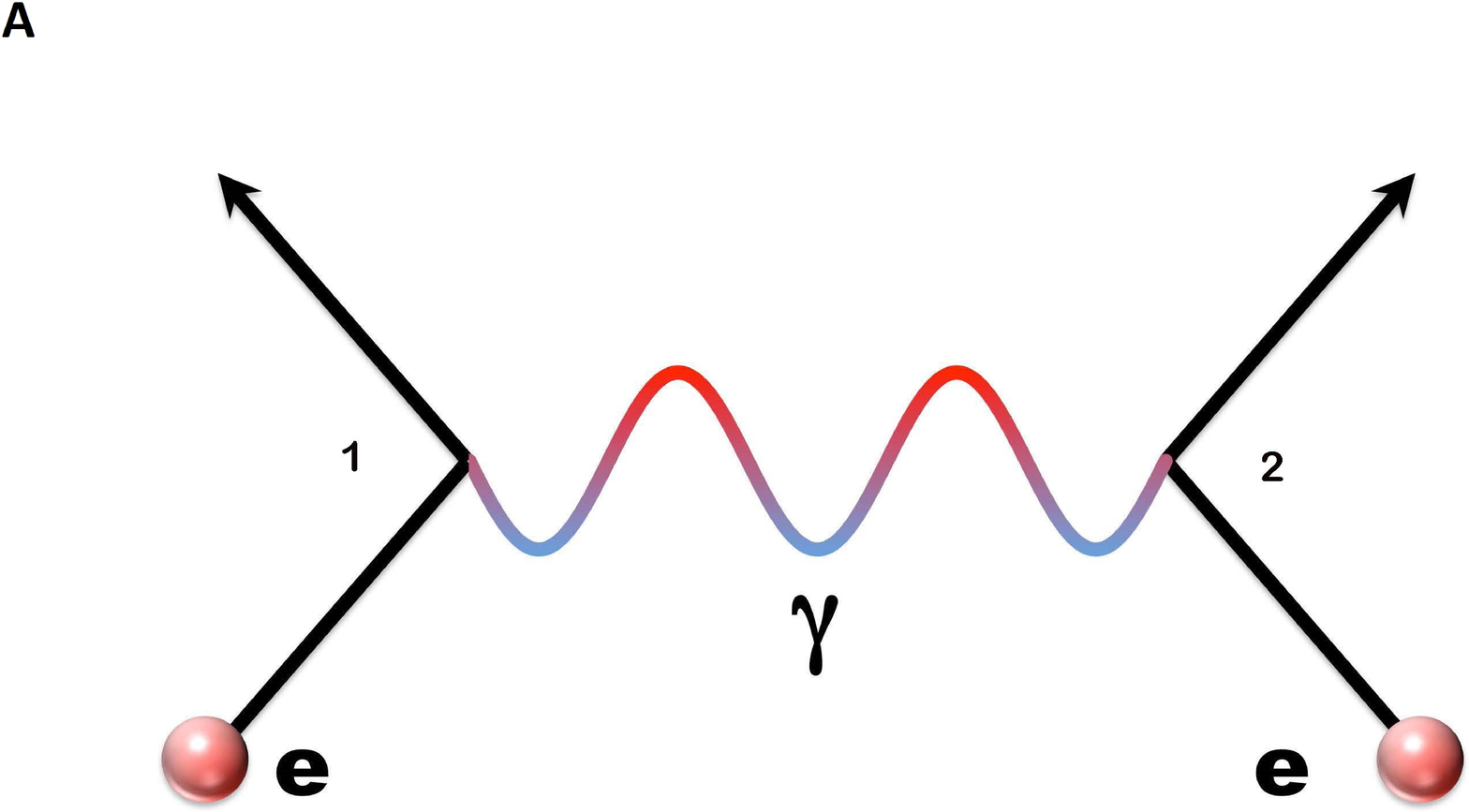}
		\label{subfigA}
		 \includegraphics[width={0.3\textwidth}]{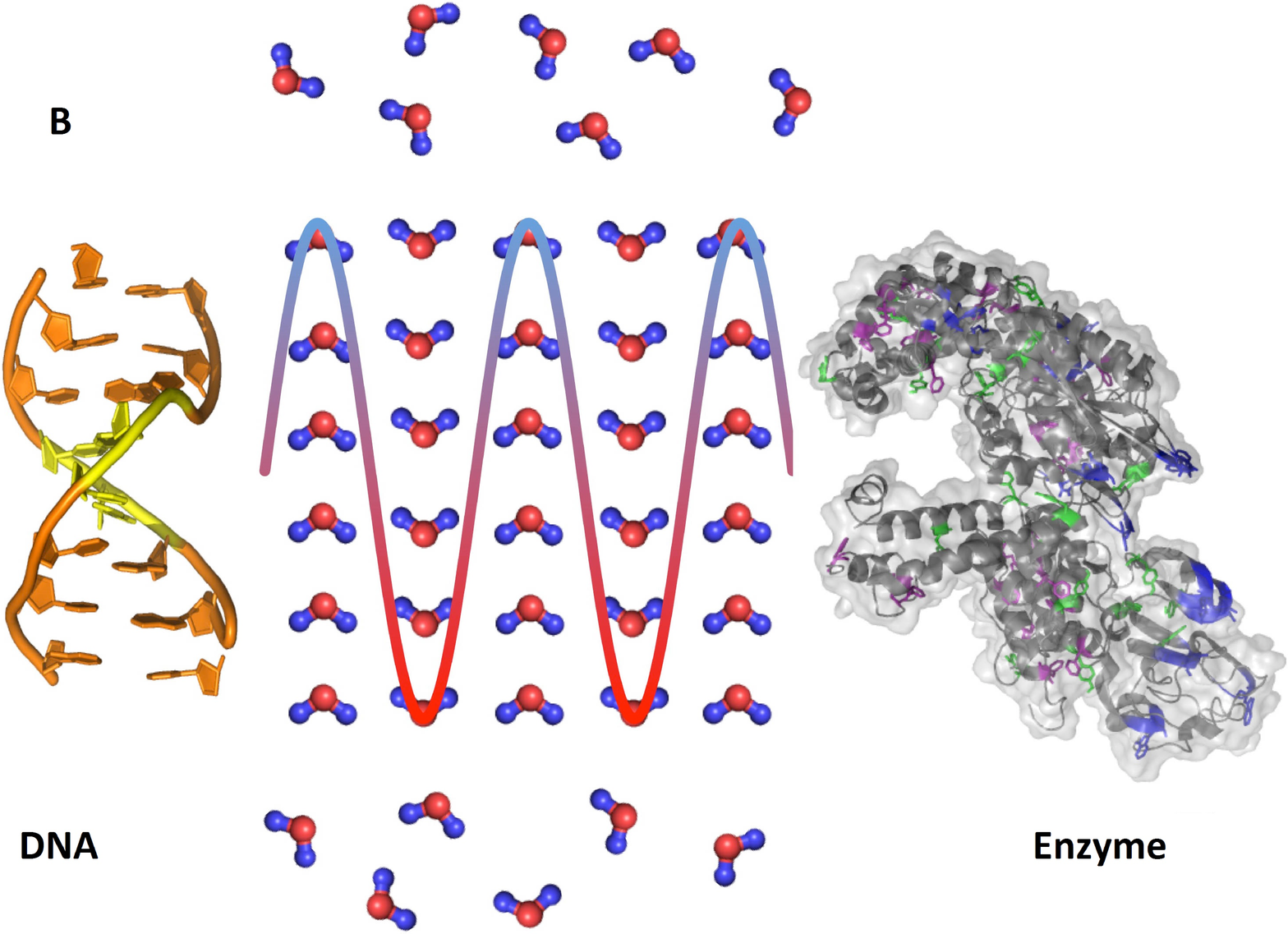}
		\label{subfigB}
\caption{\textbf{Mediating wave fields in subatomic and biological physics}. (A) In quantum field theory electron-electron interactions are mediated by photons propagating between vertex 1 and vertex 2. (B) Analogously, long-range correlations between DNA and enzymes are mediated by dipole waves in the water matrix. These renderings are schematic, neither drawn to scale nor representative of the actual orientations of water molecules. Adapted from Ref.~\cite{watermediated}.}
\label{fig:feynman}
\end{figure*}

\section{The {\it Taq}-DNA interaction and open questions
}

The enzymatic synthesis of DNA is a major subject of molecular biology, to which a large number of studies have been devoted.
The elucidations of the double-helix structure of DNA and its scaffolding by water and the hydrogen bonding of its bases in a planar configuration, directly opened the way to understanding DNA duplication. It is generally accepted that in the condition existing in living cells, DNA polymerases can operate on short open single stranded regions at  optimal temperatures lower than 40 ${}^{\circ}$C, i.e. much lower than the temperature required to melt the DNA double helix in water solutions. A peculiar exception is that of DNA  replication in thermophilic  bacteria of hot springs and in organisms living under high pressure around hot chimneys close to submarine volcanoes.

The DNA  polymerases of thermophilic bacteria have evolved to keep their structure and functionality after being exposed to heating at temperatures  close to the boiling point of water.   Kary Mullis used the polymerase ({\it Taq}) of {\it T. aquaticus}  to alternate the cycles of  heating at 95 ${}^{\circ}$C and of cooling at temperatures to allow some specific oligonucleotide primers to recognize sequences of DNA and thus amplify that DNA in a quasi-exponential manner (polymerase chain reaction or PCR) in 30 to 40 cycles.

There remains an open question: How are primers and {\it Taq} able to locate from large cellular distances and recognize short, specific DNA sequences among millions of base pairs of genomic DNA, which is a requirement to initiate DNA polymerization in PCR?
The classical explanation that Brownian  motion will permit the synchronized  meeting of the appropriate molecules is not tenable.
The highly efficient coordination of DNA-enzyme processes has been shown to confound the explanatory reach of statistical methods used to describe average regularities in biological matter\cite{Schr}. As Schr\"odinger observed for living matter in general (and his remarks  apply equally well to the case of PCR processes), the attempt to explain the high efficiency in terms of ordering generated by ``statistical mechanisms''\footnote{\cite{Schr}~(p. 47)} would be ``the classical physicist's expectation that far from being trivial, is wrong''\footnote{\cite{Schr}~(p. 19)}, and ``it needs no poetical imagination but only clear and sober scientific reflection to recognize that we are here obviously faced with events whose regular and lawful
unfolding is guided by a mechanism entirely different from the probability mechanism of physics''\footnote{\cite{Schr}~(p. 79)}.
We propose in this paper and in~\cite{DNA2a,DNA2} that a third party intervenes in the DNA-enzyme interaction processes, which is organized (coherent) water, and we show how this fits well with the general gauge theory paradigm of quantum fields~\cite{Itz,Blasone2011} (see also \cite{PRA2006,PRL1988,NuclPhys85,NuclPhys86,Alfinito:2001mm}).

Our standpoint is consistent with previous experimental work~\cite{DNA1a,DNA1,DNA2a,DNA2} that has shown that some DNAs
in water dilutions emit electromagnetic waves of low frequency (from hundreds to $\approx$ 3 kHz) which could be triggered, as in a resonance phenomenon, by an outside excitation, such as the Schumann waves~\cite{Schumann}. Experimental peaks of Schumann waves are at 7.83, 14.3, 20.8, 27.3, and 33.8 Hz; the peak near 7 Hz is the one found to be minimally active in the case discussed here (for further details see \cite{DNA1a,DNA1,DNA2a,DNA2}). The electromagnetic signals (EMS) can be amplified and recorded as a digital form. In Figure 2  the log-log plot of the EMS from the water solution of a DNA HIV-1 fragment (LTR, 194 bp) (cf. Appendix A) is reported. The decimal dilutions of original concentration of 2 ng/ml HIV-1 emitting EMS are between $10^{-6}$ and $10^{-10}$  (the first of the serial dilutions do not emit detectable signals)~\cite{DNA1a,DNA1,DNA2a,DNA2}. In the range between about 100 and 2000 Hz the signal shows fluctuations densely distributed around the linear fit with negative slope coefficient d = 0.80794, thus exhibiting scale-free, self-similar behavior in that frequency range with self-similarity (fractal) dimension d. It is known~\cite{QI,PLA2012,Systems,NCNM} that such a fractal-like self-similarity structure is the manifestation of coherent dynamics active at a microscopic level, as we indeed find in the following sections.

\begin{figure*}[htb]\label{HIV}
\centering
\includegraphics[width=0.5\textwidth]{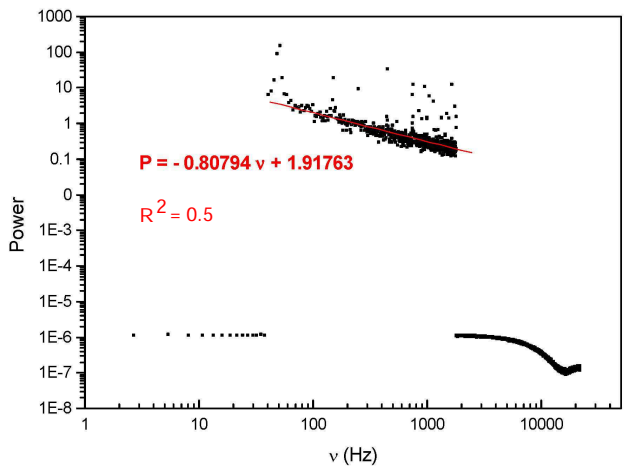}
\caption{\textbf{Logarithmic plot of power  vs frequency of EMS emitted from water solution of DNA HIV-1 long terminal repeat (LTR 194 base pairs).} The decimal dilutions of  original concentration of 2 ng/ml HIV-1 emitting EMS  are between $10^{-6}$ and $10^{-10}$~\cite{DNA1a,DNA1,DNA2a,DNA2}. The signal analysis is limited to the frequency range between about 100 and 2000 Hz. The signal in the frequencies below $\approx$ 40 Hz and above $\approx$ 2000 Hz (dotted lines) does not show self-similar behavior.}
\end{figure*}

This first step can be followed in different conditions of space and time by a second step, consisting in transformation  of the signal into an analog form and  sending  its electric  vector to a solenoid which will generate a magnetic field in a test tube of water. Specific water nanostructures (domains of coherently oscillating molecular dipoles) will be  formed (``signalized'' water).
Then the electromagnetic image of the initial DNA is ``read'' by the {\it Taq} polymerase in the  presence of primers and nucleotide triphosphates,  generating and amplifying DNA molecules identical to the initial DNA from {\it Borrelia burgdorferi} or the LTR of HIV-1.

An interesting  variation of the second step is to replace the test tube of water by a flask  of living cultured cells.
If  the  cells  were originated from malignant tumor  or isolated {\it in vitro} as continuous  ``immortalized'' cell lines, after several days of exposure  to EMS of a foreign bacterial DNA amplicon (i.e. a fragment of DNA or RNA that is source and/or product of amplification or replication event), they  synthesize that DNA which can be detected  by usual PCR. At the same time the growth of the exposed cells is inhibited and finally they die.

One interpretation of this surprising result is that such cells possess  an enzymatic system analogous to the {\it Taq} polymerase, i.e., able to read the DNA information contained  by the nanostructures induced in  the water  of the cells.  This enzymatic system analogous to the {\it Taq} polymerase synthesizes the  bacterial  DNA in the presence of a pool of nucleotide triphosphates, but without the need of  specific primers. This pool in multiplying cells may be smaller than the one artificially provided in water. This difference may explain the delayed effect of EMS carrying the bacterial DNA in living cells. No effect was observed if the exposure to EMS was restricted to one or two days. The origin and location of the cellular enzymatic system has not been determined, although it is sensitive to inhibitors of mitochondria. This enzymatic system may be a modified cellular polymerase or an enzyme of bacterial origin close to it.
It  should be  noted that the reconstruction of foreign bacterial DNA either by the water of a test tube or in the interior of cells has been attained only if this DNA was extracted from pathogenic bacteria or virus and emitted recordable EMS~\cite{DNA1a,DNA1,DNA2a,DNA2}. The details of this anti-tumor effect of EMS of  bacterial origin will be the object of another paper.

These previous results fit very well with the two vertices hypothesis developed  below. If we generalize it to the PCR working on any cellular DNA, we  have then to make the assumption that the EMS are too weak to be detected and amplified under  natural  conditions, but are still sufficient to trigger a coherent dynamics in the water. In the following we schematize  the two-vertex interaction according to the analysis we present in Section 4 (see Appendix A for the operative protocol for DNA transduction in water):\\

$\bullet$ First vertex : DNA-water interaction vertex:\\

The DNA-water EMS is detected, amplified and digitized ${\bf \rar}$ computer recording.

${\bf \rar}$ The recorded EMS is used for the\\

$\bullet$ Second vertex : \{signalized water\}-{\it Taq} interaction vertex:\\

The EMS is converted to analog EMS

$\rar$ played in water where the initial DNA-water em image is reproduced

(signalized water, coherent nanostructures) $\rar$ \{signalized water\}-{\it Taq} interaction vertex

$\rar$ 40 cycles of PCR $\rar$ DNA amplification.\\

When the analog EMS is played in cells $\rar$ nanostructures in cells

$\rar$ DNA $\rar$ cell death

See Refs.~\cite{DNA1a,DNA1,DNA2a,DNA2} for more details on the detection of EMS from DNA water solutions and DNA transduction in water and in cells. In Section 5 we discuss a few characterizing properties of the interactions at the two vertices on the basis of results presented in Sections 3 and 4.

\section{Collective dipole behavior in DNA and enzyme molecules}

The electromagnetic radiative properties of DNA and enzyme macromolecules depend on the characterization of their electric dipoles. Therefore we first consider these molecular dipole configurations, summarizing the analysis presented in Refs.~\cite{watermediated} and ~\cite{Kurian}.

The molecular structure of DNA base pairs and enzyme amino acids, as well as many other biomolecules, presents aromatic rings containing conjugate planar systems with $\pi$ electrons delocalized across the structure. A typical example of such rings is benzene. Such an electronic configuration is at the origin of the formation of dipolar arrangements, which in turn can produce non-trivial electromagnetic couplings among biomolecules~\cite{Ambrosetti2016}. It is also at the origin of the stability and low reactivity in aromatic compounds.

In the DNA a special role is played by the symmetry of the molecular distribution about its helical axis. The dipole-dipole interaction is given by
\begin{equation} \label{eq:dip-dip}
V_s^{int}
= \frac{1}{4\pi\epsilon_0 d^3}
\left[\gv{\pi}_s \cdot \gv{\pi}_{s+1}- \frac{3(\gv{\pi}_s \cdot \mathbf{d})(\gv{\pi}_{s+1} \cdot \mathbf{d})}{d^2} \right],
\end{equation}
with $\mathbf{d} = d\hat{\mathbf{z}}$ connecting the centers of nearest-neighbor base-pair dipoles $\gv{\pi}_s = Q\mathbf{r}_s$ and $\gv{\pi}_{s+1} = Q\mathbf{r}_{s+1}$.  Anisotropies are accounted for by the numerical tensor elements (cf. Table 1 in Appendix B) $\alpha_{ii}$, which are generally determined to within five percent agreement from perturbation theory \cite{mcweeny1962perturbation}, simulation \cite{papadopoulos1988polarisability}, and experiment \cite{basch1989electrical}.

By following the analysis presented in \cite{Kurian}, one may introduce the normal-mode lowering operator $a_{s, j}$
and its conjugate raising operator $a_{s,j}^\dagger$ for the $s^{th}$ normal mode of the collective electronic oscillations, with
$s=0,1,\dots,N-1$, $j=xy, z$ (see Appendix B for their explicit expressions). The system Hamiltonian takes then the standard diagonalized form
\begin{equation} \label{eq:numop}
H_{DNA}=\sum_j H_j= \sum_j \sum^{N-1}_{s=0}\hbar\Omega_{s,j} \left (a_{s,j}^\dagger a_{s,j} +\frac{1}{2} \right).
\end{equation}
The eigenstates of $H_j$ are given by $\ket{\psi_{s,j}}=a_{s, j}^\dagger \ket{0}$. In particular we examine the so-called ``zero-point'' modes, on which the number operator $n_{s,j}=a^{\dagger}_{s,j} a_{s,j}$ in Eq.~(\ref{eq:numop}) acts to produce a zero eigenvalue, because these are most easily excited by the free energy changes due to enzyme clamping. They are collective normal modes of the DNA system considered in our model framework. Because these zero-point oscillations are normal modes, a four-base-pair sequence will produce four frequencies of coherent (phase-synchronized) oscillation in the longitudinal direction (see Appendix B for $\Omega_{s,z}$ in the four-bp case when DNA sequence homogeneity is assumed).

It is noteworthy that in order to break a single phosphodiester bond in the DNA helix backbone, the required energy \textit{in vivo} is $\varepsilon_{P-O} \simeq 0.23$ eV \cite{dickson2000determination}, which is less than two percent of the energy required to ionize the hydrogen atom, but about ten times the physiological thermal energy ($k_B T$), and comparable with the quantum of biological energy released during nucleotide triphosphate (e.g., ATP) hydrolysis. The meaning of this is that the bonds of the DNA backbone are not so tight as to be unmodifiable, but they are strong enough to resist thermal degradation. As shown in Ref.~\cite{Kurian}, at the standard inter-base-pair spacing of $3.4\,$\r{A}, the ground state longitudinal oscillation (a zero-point mode) for a six-bp DNA sequence is calculated to within $0.5\%$ of $2\varepsilon_{P-O}$, the energy required for double-strand breakage.

\textit{Taq} represents a broad class of enzymes, called DNA polymerases, which replicate DNA strands with high efficiency.
It does not recognize specific sequences, but it uses a DNA clamp, which is an accessory multimeric protein encircling the DNA double helix as the polymerase adds nucleotides to the growing strand. The clamp increases the number of nucleotides the polymerase can add to the growing strand per binding event, up to 1,000-fold, aided by a layer of water molecules in the central pore of the clamp between the DNA and the protein surface. Due to the toroidal shape of the assembled multimer, the clamp cannot dissociate from the template strand without also dissociating into monomers.

Amino acids are the building blocks of protein and enzyme systems. The most polarizable amino acids are termed \textit{aromatic}, and as with DNA base pairs, they induce dipole interactions between their constituent ring structures. The most aromatic amino acids (tryptophan, tyrosine, and phenylalanine) are also the most hydrophobic. It has been argued that London force dipoles in such intra-protein hydrophobic pockets could couple together and oscillate coherently, thus generating a radiative field.

\begin{figure}
\centering
		\includegraphics[width={0.5\textwidth}]{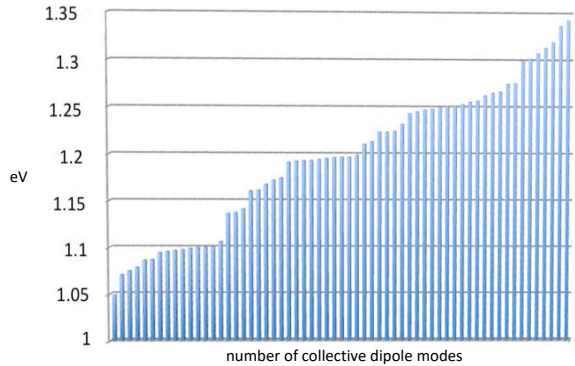} \label{subfigB}
\caption{\textbf{Collective dipole oscillations in aromatic amino acid networks of  \textit{Taq} DNA polymerase.} Collective normal-mode solutions to networks of aromatic induced dipoles in \textit{Taq} are within the energy range of the collective dipole modes of DNA bounded by its relevant protein clamps. In the diagonalized form of the coupled harmonic oscillator Hamiltonian, the number operator $n_{s} = a_{s}^\dagger a_{s}$ acts on these states to produce a zero eigenvalue, analogously to Eq. (\ref{eq:numop}) for DNA. They are identified with zero-point modes of the ground state of the enzyme aromatic network. Data for the collective modes (arranged along the abscissa axis according to increasing energy) are presented in units of eV on the ordinate axis. Adapted from Ref.~\cite{watermediated}.}
\label{enzcohmodes}
\end{figure}

In regards to the spacing between aromatic amino acids in protein, when considering tryptophan, tyrosine, and phenylalanine, the spacings are as close as ~5 \r{A} in the enzymes of interest, comparable to the inter-dipole spacing of DNA (3.4 \r{A}).
We consider the pairwise interaction potential $V_{s,t}^{int}$ between each amino acid electric dipole in the enzyme's aromatic network. $V_{s,t}^{int}$ has a similar expression to the one for DNA considered above (cf. Eq.~(\ref{eq:dip-dip})), but without helical symmetry and including pairwise, not just nearest-neighbor, aromatic dipole couplings. The enzyme collective modes, like the DNA modes, are dipole fluctuation modes and can be used to construct the creation and annihilation operators for the radiative electromagnetic field quanta acting on the water molecules. Collective modes for \textit{Taq} are presented in Figure \ref{enzcohmodes}.  The average energy for these \textit{Taq} modes is 1.2 eV.

Though the aromatic amino acids of proteins typically absorb in the UV, around 4 - 5 eV, this observation does not discount coherent oscillations of induced dipoles at longer wavelengths in protein aromatic networks due to the dipolar correlations described in Refs. \cite{watermediated, Kurian}. It is quite remarkable that the enzyme collective dipole modes displayed in Figure \ref{enzcohmodes} fall within the spectrum of DNA collective modes described in full detail in Ref. \cite{Kurian}.

We now turn to study the coupling between the dipole field of the molecular water bath and the radiative dipole fields of the DNA and {\it Taq} enzyme. The biochemical activity between DNA and enzyme in PCR is conditioned and thus established by these long-range correlations.

\section{DNA and {\it Taq} coupling to the water dipole field}

In this section we show that the electromagnetic pattern of the radiative dipole fields generated by DNA and enzyme induces in the water environment dipole wave fields which mediate the DNA-enzyme interaction. We study the DNA-water interaction ``vertex'', and the enzyme-water interaction ``vertex'' (Figure 1). Our analysis applies to the PCR processes where the DNA template is used and to PCR processes where signalization of the water replaces the use of the DNA template as described in Section 2 (and in Appendix A).  We will neglect the static dipole-dipole interactions between DNA and water dipoles and between enzyme and water dipoles, focusing our attention instead on the radiative dipole field.

Consider a spatially uniform distribution of $N$ molecules of liquid water in a volume $V$ of linear  size $\ell$ at a non-vanishing temperature $T$. In the notation of Ref.~\cite{watermediated} and \cite{PRA2006}, denote the water molecule dipole vector by $\mathbf{d}_e$, with magnitude  $|\mathbf{d}_e | =2\,e\,d_{e}^{2}$,  $d_{e} \simeq 0.2\,$\r{A} \cite{Franks} and $e$ the electron charge. We will use natural units, where $\hbar = 1 = c$. In the absence of DNA and enzyme molecules, the electric dipoles of water molecules are arbitrarily oriented and there is no preferred orientation direction; this is expressed by saying that the system of water molecules is invariant under dipole rotations. Such a dipole rotational symmetry is broken by the interaction with the dipole radiative fields of DNA and enzyme molecules.  Our aim is to study the collective rotational transitions in the water dipole field induced by such radiative interactions \cite{PRL1988,Tsenkova1,Tsenkova2,Tsenkova3,MarkHenry}.

Let $\chi ({\bf x},t)$ be the complex dipole wave field collectively describing the system of $N$ water molecules in the unit volume $V$ (cf. Appendix C). In the absence of DNA and enzyme, the assumption that molecule density is spatially uniform implies that the only relevant variables are the angular ones. Therefore $\chi ({\bf x},t)$ may be expanded in terms of spherical harmonics on the unit
sphere as $\chi ({\bf x},t) = \sum_{l,m} \alpha_{l,m}(t)Y^{m}_{l}(\theta, \phi)$.

For our purposes it is enough \cite{watermediated,PRA2006} to consider the expansion in the four lowest-energy levels $(l,m) = (0,0)$ and $(1,m)$ for $m = 0, \pm 1$, effectively setting $\alpha_{l,m}(t) = 0$ for $l \neq 0,~ 1$. At thermal equilibrium, the populations of these levels are given by $N |\alpha_{l,m}(t)|^{2}$ and follow the Boltzmann distribution. We will show that the populations of the $l=0$ and $l=1$ levels change once we consider the presence of DNA and enzyme in water.

It is useful to introduce the notations
\bea \nonumber \al_{0,0}(t) &\equiv& a_{0}(t) \equiv
A_{0}(t)e^{i\de_{0}(t)},\\
 \al_{1,m}(t) &\equiv&  A_{1}(t)e^{i\de_{1,m}(t)}e^{-i\om_{0}t} \equiv a_{1,m}(t)
e^{-i\om_{0}t}. \lab{9} \eea
$A_{0}(t)$, $A_{1}(t)$, $\de_{0}(t)$, and $\de_{1,m}(t)$ are real
quantities and  $\om_{0} \equiv {1}/{I}$ (in natural units). We adopt the value $I = 1.95943\times 10^{-40}$ g$\cdot$cm$^2$ as the average moment of inertia of the water molecule (Appendix C); $\om_{0}$ is the eigenvalue of ${{\bf L}^{2}}/{2I}$ on the state $(1,m)$, with ${\bf L}$ denoting the total angular momentum for the molecule. Indeed, ${l(l+1)}/{2I} = {1}/{I}$. We restrict our discussion to the resonant radiative em modes, i.e., those for which $\om_{0} \equiv k = {2\pi}/{\la}$, and we use the dipole approximation, where $\exp(i{\bf k}\cdot {\bf x}) \approx 1$, since we are interested in the collective behavior of the system of $N$ molecules. The system linear size $\ell$ is then constrained to be $\ell \leq \la = {2\pi}/{\om_{0}}$;  $\om_{0}$ thus provides a relevant scale for the system. For water, ${2\pi}/{\om_{0}} \approx 1774 \,(\text{eV})^{-1} \approx 3.5 \times 10^{-2}\,$cm (in $\hbar = 1 =c$ units).

The dipole rotational invariance implies that the amplitude of $\alpha_{1,m}(t)$ does not depend on $m$, or, stated in different words, the three levels $(1,m)$ for $m = 0, \pm 1$ are in the average equally populated under normal conditions, thus $\sum_{m} ~|\al_{1,m}(t)|^{2} =  3 ~|a_{1}(t)|^{2}$, with $a_{1}(t) \equiv a_{1,m}(t)$ for any $m = 0,\pm1$.
Dipole rotational invariance also implies that the time average of the polarization $P_{{\bf n}}$ along any direction ${\bf n}$ vanishes. In fact, considering a vector ${\bf n}$ parallel to the ${\bf z}$ axis in an arbitrary reference frame, one can show that the time average of $P_{{\bf n}}$ is zero, as it must be:
\bea  P_{{\bf n}} = \int d\Om \chi^{*} ({\bf x},t)({\bf
x} \cdot {\bf n})\chi ({\bf x},t)
= \frac{2}{\sqrt{3}}A_{0}(t)A_{1}(t)\cos(\om -\om_{0})t ~, \lab{8}
\eea
where $\om t \equiv \de_{1,0}(t) - \de_{0}(t)$. In such a general setting, the initial conditions at $t = 0$ can be written as
\be \lab{4f} |a_{0}(0)|^{2} = \cos^{2} \theta_{0},~~\qquad
|a_{1}(0)|^{2} = \frac{1}{3}\sin^{2} \theta_{0}. 
\ee
by excluding of course the values of $\theta_{0}$ corresponding to the  completely filled state (0,0) ($\theta_{0} = n\pi$, n integer) or completely empty ($\theta_{0} = (2n + 1){\pi}/{2}$, n integer), which are in fact physically unrealistic states.  Interestingly, one finds that the dynamics imposes in a self-consistent way the lower bound for the parameter $\theta_{0}$.  Also note that the equipartition of the field modes among the four levels $(0,0)$ and $(1,m)$ given by the Boltzmann distribution for $k_{B} T \gg E(k)$, with $k_{B}$ the Boltzmann  constant and $E(k)$ the energy of the field modes, corresponds to $\theta_{0} = {\pi}/{3}$.

Let us now consider how the above scenario changes when the DNA-water interaction is present. We need to consider the effects due to the electromagnetic field generated by the dipole vibrational modes of the DNA, discussed in Section 3, and of the water molecules. Let us denote it by $u_{m}(t) \equiv U (t) e^{i\varphi_{m}(t)}$, with $U (t)$ and $\varphi_{m}(t)$ real quantities. One can show~\cite{PRL1988,PRA2006,MarkHenry} that the amplitude and the phases do not depend on $m$, so that we may write $|u(t)| \equiv |u_{m}(t)|$, $\varphi \equiv \varphi_{m}$, $\de_{1}(t) \equiv \de_{1,m}(t)$.
We consider the coupling of $u(t)$ with water molecules in the transition $(0,0) \leftrightarrow (1,m)$.

The field equations for a water molecule in the field $u(t)$, for which it is assumed the transversality condition ${\bf k} \cdot {\gv \ep}_{r} = 0$, with ${\gv \ep}_{r}$ the polarization vector of the em mode,  are given in Refs. \cite{Knight,Heitler} and for brevity we do not report them here (see Refs. \cite{watermediated} and \cite{PRL1988}).
An inspection of these equations shows that the strength of the coupling between the em modes and the molecular levels is given by $\Om = {4ed_{e}}\sqrt{{N}/({6\om_{0}V})}~\om_{0} \equiv G~\om_{0}$. Under standard conditions, for pure water, $G \simeq 13$. The coupling $\Om$ scales with the density $\sqrt{\rho} \equiv \frac{N}{V}$ and thus varies with temperature and pressure.
The initial conditions (\ref{4f}) imply that at $t=0$ we have $|u(0)|^{2} = 0$.

We remark that the analysis of the field equations shows that $a_{0}(t) = 0$, for any $t$, is an instability point so that the system
spontaneously runs away from $a_{0}(t) = 0$ (which also justifies and is consistent with our assumption made on physical grounds (see the comments following Eq. (\ref{4f}))). It can be shown~\cite{PRA2006,PRL1988} that the spontaneous breakdown of the
phase symmetry implies that the system runs away from the state with $u (t) = 0$ for any $t$, which excludes the values $\theta_{0} \leq {\pi}/{4}$.
Dynamical consistency then requires that  $|u(t)|^{2}  = - ({1}/{3}) \cos 2\theta_{0} \equiv v^{2}(\theta_{0})$, with  $\theta_{0} > {\pi}/{4}$, which is time-independent. The motion equations also lead to
\bea \lab{24a}  \dot{U}(t) &=& 2\Om A_{0}(t)A_{1}(t) \cos \al (t) ~\\
\lab{24b} \dot{\varphi}(t) &=& 2\Om
\frac{A_{0}(t)A_{1}(t)}{U(t)}\sin \al (t), \eea
where $\al = \de_{1}(t) - \de_{0}(t) - \varphi(t)$. From these equations we see that a time-independent
amplitude $\overline{U}$ exists, i.e., $\dot{U}(t) = 0$,  if and only if the phase-locking relation
\be \lab{25} \al = \de_{1}(t) - \de_{0}(t) - \varphi (t) =
\frac{\pi}{2} \ee
is satisfied. Eq.~(\ref{25}) gives
\be \lab{26}  \dot{\varphi}(t) =  \dot{\de_{1}}(t) -
\dot{\de_{0}}(t) = \om,\ee
with $\om$ introduced in Eq.~(\ref{8}). This dynamical regime is the so-called ``limit cycle''. By using the normalization condition and the constants of motion, one can obtain
\be \lab{26g} \overline{A}_{0}^{2} - \overline{A}_{1}^{2} =
\cos^{2} \theta_{0} - \frac{1}{3}\sin^{2} \theta_{0} + 2\overline{U}^{2} \neq 0, \ee
to be compared with $A_{0}^{2}(t) - A_{1}^{2}(t) \approx 0$ at thermal equilibrium in the absence of collective dynamics. The spontaneous breakdown of the phase symmetry produces as a dynamical effect a coherent em field pattern (``limit cycle'' state): the interaction of the em field with the molecular water dipoles is characterized by the ``in-phase" (coherent) dynamics expressed by the phase-locking Eq. (\ref{25}).  Eq. (\ref{26}) means that any change in time of the difference between the phases of the amplitudes $a_{1}(t)$ and $a_{0}(t)$  is compensated by the change of phase of the em field. The coherence between the matter field  and the em radiative dipole field guarantees the gauge invariance of the theory. The formal expression of the collective dynamics is thus expressed by Eqs. (\ref{25}) and (\ref{26}). In summary, the combined system of DNA and water's radiative em field evolves away from the initial symmetric ground state to the asymmetric ground state $|u(t)|^{2}  \neq 0$.

We consider now the enzyme radiative dipole field interacting with the em field pattern generated by the DNA-water interaction.
We add to the system energy the interaction term ${\cal H} = - {\bf d_{e}} \cdot {\bf E}$, with ${\bf d_{e}}$ the electric dipole moment for water, and ${\bf E}$, assumed parallel to the ${\bf z}$ axis, is the enzyme radiative electric field.
The Hamiltonian ${\cal H}$ can be written in the form of a Jaynes-Cummings-like Hamiltonian for large $N$~\cite{Knight}:
\be \lab{JC} {\cal H} = \hbar \sqrt{N}\ga (a^{\dag} S^{-} + ~a S^{+}).
\ee
In Eq.~(\ref{JC})  ~$\ga$ is the coupling constant proportional to the matrix element of the molecular dipole moment  and to the inverse of the volume square root $V^{-1/2}$; $a^\dagger$ and $a$ are the creation and annihilation operators for the electric field ${\bf E}$, $S^{\pm}$ denote the creation and annihilation operators of the dipole wave modes of the field pattern imaging the DNA-water interaction.

One can show that, by using the expansion of $\chi ({\bf x},t)$ in terms of  spherical harmonics, ${\cal H}$ induces the mixing:
$Y^{0}_{0} \rightarrow Y^{0}_{0} \cos
\tau + Y^{0}_{1} \sin \tau$ and $Y^{0}_{1} \rightarrow - Y^{0}_{0}
\sin \tau + Y^{0}_{1} \cos \tau$, with
\be \lab{4pa} \tan \tau = \frac{\om_{0} - \sqrt{\om_{0}^{2} + 4
{\cal H}^{2}}}{2{\cal H}}. \end{equation}
The polarization $P_{{\bf n}}$ becomes now
\bea  P_{{\bf n}} = \frac{1}{\sqrt{3}} (\overline{A}_{0}^{2} - \overline{A}_{1}^{2})
\sin 2\tau +  \frac{2}{\sqrt{3}} \overline{A}_{0}^{2}\overline{A}_{1}^{2} \cos 2\tau  \cos [(\om - \om_{\cal H})t],  \lab{4pb}  \eea
where $\om_{\cal H} \equiv \sqrt{\om_{0}^{2} + 4 {\cal H}^{2}}$ and $\overline{A}_{0}^{2} - \overline{A}_{1}^{2}$ has the limit cycle non-zero value (cf. Eq.~(\ref{26g})).  The time averaged polarization is $\overline{P_{{\bf n}}} = (1/\sqrt{3}) (\overline{A}_{0}^{2} - \overline{A}_{1}^{2}) \sin 2\tau$. It is non-vanishing provided that $\tau \neq 0$ and that $(\overline{A}_{0}^{2} - \overline{A}_{1}^{2})\neq 0$.
The former condition ($\tau \neq 0$) is realized due to the enzyme electric dipole field, as we have seen above. The latter condition ($\overline{A}_{0}^{2} - \overline{A}_{1}^{2} \neq 0$) is realized in the water coherent domains imaging the DNA em radiative field. $\overline{P_{{\bf n}}}$ therefore depends on both the DNA radiative field and the enzyme field, and it is zero if one or both fields are absent. It represents their long-range dipole wave correlation.

The physical meaning of the anthropomorphic expression ``the enzyme sees the DNA'' is that the enzyme couples to the collective  wave mode generated by the radiative em dipole interaction between DNA and water. Such a wave mode accounts for local dependence and space-time distributions of couplings, amplitudes, and frequencies along the DNA molecular chain. The resulting wave mode envelope describes the em dynamical image of the DNA in the surrounding water environment. This water-mediated DNA em image is what the enzyme actually ``sees''. The converse can be said from the standpoint of the DNA, which sees the enzyme em image. The action of reciprocal ``seeing'' is described by the exchange of the dipole wave quanta of the collective dynamics induced by DNA and enzyme in the surrounding water dipole field, as derived above. This suggests that in widely used PCR processes for the amplification of DNA sequences, what is really crucial for their successful occurrence is the formation of the DNA em image in the water and the possibility for \textit{Taq} DNA polymerase to recognize such an image by dipole wave quanta exchange as described above. This may be helpful in setting the stoichiometric parameters in the PCR preparation, and a study in this direction is in our plans. Our analysis 
is also consistent with the observation that PCR amplification of DNA is possible through the production of an em image of DNA in pure water by its exposure to the recorded dipole wave radiation emitted by the DNA-water interaction vertex.

\section{Discussion and Concluding Remarks}

In this paper we have studied the role played by the water radiative dipole wave bridging the interaction between DNA and enzyme macromolecules in the process of PCR. The bridge is realized by the collective long-range correlation triggered in the water by the presence of the dipole structures of DNA, on the one hand, and the enzyme macromolecule, on the other. These two ``interaction vertices'' have been studied in the frame of the conventional gauge field theory paradigm, stating that any interaction between two systems is mediated by the propagation of a correlation field or quantum, such as, for instance, the photon exchanged between interacting electric charges in quantum electrodynamics.

As far as we know, our study provides the first dynamical description of the conventional PCR process, which is traditionally analyzed in the literature by means of stoichiometric and thermodynamics-based methods. Our study does not contradict or substitute traditional biochemical studies. Their description of average regularities in biological systems in terms of purely statistical methods remains valid. What our study provides is the {\it dynamical} description of the long-range em correlation modes orchestrating the PCR process, thus accounting for its remarkable efficiency, space-time ordering, and diverse time scales that would be unattainable by otherwise fully stochastic molecular activities. As observed in the previous section, our discussion shows that in PCR processes what is essential for successful DNA amplification is the formation of a coherent dipole wave pattern representing the  em image of DNA in  water and the capability  of \textit{Taq} DNA polymerase to recognize such an image. The general process is described by the analysis of two interaction ``vertices'': the interaction vertex DNA-\{water dipole wave\}, on the one hand, and the interaction vertex \{water dipole wave\}-\textit{Taq}, on the other.

In the standard DNA-\textit{Taq} interaction, this image recognition would aid in the location of the DNA substrate by the polymerase, resulting finally in conformational changes, exclusion of water molecules, and direct polymerase binding. Direct binding would alter \textit{Taq}'s conformational and vibrational state, effectively ``reading'' the DNA and catalyzing chemical reactions to add appropriate nucleotides complementary to the template sequence. Thus, the radiation field from the DNA vertex induces conformational change in the \textit{Taq} polymerase at the other vertex, just like in the conventional description where the chemical DNA template induces direct polymerase binding. What \textit{Taq} really recognizes and ``reads'' is the em image of DNA, more specifically the em signals emerging from the collective molecular electrodynamics of the DNA template's constituent charges. It is consistent with our analysis that the em image of DNA may be thus produced by ``signalizing'' pure water through the coherent dipole wave ``signals'' emitted from the interaction vertex DNA-\{water dipole wave\}, according to the protocol reported in Appendix A.

Of course, a crucial role is played by the distributions of charges and dipoles, and any other conformational and dynamical details of the DNA and \textit{Taq} molecular structures. Moreover, a fundamental role is played by the (fractal-like) self-similarity (Figure 2) of the EMS emitted by the DNA-\{water dipole wave\} interaction vertex, which has been shown~\cite{QI,PLA2012,Systems,NCNM} to be the manifestation of coherent microscopic dynamics, as indeed our discussion in this paper has suggested. These (fractal-like) self-similarity properties of the EMS in turn induce self-similar organization in the irradiated water. This is often overlooked in chemical spectral analysis. Not only is the spectral composition of the EMS important, but much more relevant to our purposes are these self-similarity (coherence) properties. In some sense, by reverting the argument, one might also say that the observed amplification of DNA in PCR processes appears to provide a ``measurement device'' for these self-similar structures produced in signalized water.

In order to further analyze the physical properties of the two interaction vertices considered above, we observe that the characteristic time scales of the process are controlled by the radiative dipole field amplitude and, remarkably,  they are much shorter than the ones of thermal noise. As a consequence, the collective dynamics is protected by an energy gap. This is better clarified by observing that the factor $\sqrt{N}$ multiplying the coupling $\ga$ (cf. Eq.~(\ref{JC})) implies that for large $N$ the collective interaction energy scale is larger by the factor $\sqrt{N}$ than the microscopic energy fluctuations of short-range molecular interactions, thus providing a protective gap against thermalization for the long-range quantum correlations. Conversely, the collective interaction time scale is shorter by the factor $1/\!\sqrt{N}$ than the time scale of the short-range interactions among the molecules. In a similar way, the multiplying factor $\sqrt{N}$ produces the enhancement of the coupling $G$ (cf. Eq.~(\ref{4f})). These remarks may account for the stability and efficiency of the PCR process against the high temperatures (up to 95 ${}^{\circ}$C) of the various thermal cycles, otherwise inexplicable in a purely statistical approach based on molecular kinematics.
These kinds of enhancement phenomena for the coupling constants are well known to occur in superradiance and cooperative two-level systems \cite{Celardo1, Celardo2}. As a result, the macroscopic stability of the system follows since it is protected against quantum fluctuations in the microscopic short-range dynamics, and, for sufficiently large $N$, against thermal fluctuations, unless $k_{B} T$ is of the same order or larger than the height of the protective energy gap.

We have seen that the spontaneous breakdown of the phase symmetry produces as a dynamical effect a coherent em field pattern whose interaction with the molecular water dipoles is characterized by the phase-locking of Eqs.~(\ref{25}) and (\ref{26}) (``in-phase" (coherent) dynamics). One can show (see e.g. Appendix C in Ref. \cite{watermediated}; also Ref.~\cite{PRA2006}) that the system ground state is a coherent condensate of the quanta of the long-range dipole waves  (the Nambu-Goldstone boson quanta). In the QFT formalism, gauge invariance requires fixing the phase of the matter field and a specific gauge function has to be selected. One thus introduces the covariant derivative $D_{\mu} = \pa_{\mu} - igA_{\mu}$ and the variations in the phase of the matter field lead to the pure gauge $A_{\mu} = \pa_{\mu} \varphi$. Since partial derivatives with respect to time and space coordinates can be interchanged for regular (continuous and differentiable) functions, when $\varphi({\bf x},t)$ is a regular function, then we obtain vanishing electric and magnetic fields, ${\bf E}=-\frac{\pa {\bf A}}{\pa t} +
{\bf \nabla} A_{0}= (-\frac{\pa }{\pa t}{\bf \nabla} +{\bf \nabla}
\frac{\pa }{\pa t})\varphi  = 0$, and ${\bf B}={\bf \nabla} \times {\bf A} = {\bf \nabla}
\times {\bf \nabla} \varphi= 0$. This means that in order to obtain non-vanishing values for the fields {\bf E} and  {\bf B} in a coherent region,  $\varphi({\bf x},t)$ must exhibit a divergence or topological singularity within that
region~\cite{NuclPhys85,NuclPhys86,Alfinito:2001mm}. This is indeed observed in condensed matter physics. For example, in type II superconductors,  a vortex core is penetrated by the lines of a quantized magnetic flux. Such a result is particularly relevant in our analysis, since it describes the dynamic formation of observed topologically non-trivial conformations of DNA-enzyme ``clamping'': in the $Taq$ DNA clamp, a protein multimeric structure encircles the DNA double helix, like encircling the core of a vortex, formed indeed by a layer of water molecules in the central pore of the clamp between the DNA and the protein surface. A similar ``clamping'' is observed in the DNA endonuclease catalytic activity of the \textit{Eco}RI enzyme~\cite{watermediated,Kurian}.

Let us close by observing that our analysis may contribute toward answering the question of why divalent cations like Mg$^{2+}$ play such a relevant role in the precise control of enzyme behavior for optimum biological functions, which is actually an unsolved problem in biology. We observe that the strict dependence on the pH and Mg$^{2+}$ concentration in enzyme catalytic activity suggests that these variations propagating through the buffer solution may  produce  local perturbations in the electromagnetic environment. In PCR this may produce  deformation of the electromagnetic image of DNA to the point that it is no more recognizable by the polymerase. This might thus account for the fact that  \textit{Taq} is extremely dependent on hydrogen and magnesium ion concentration, and that the success of the PCR depends critically on these concentrations. Divalent cations, in particular, may therefore enhance or maintain or critically deform the long-range dipole wave fields in water solution. This is supported by recent experimental work that shows that ions can influence up to $10^6$ water molecules, which extend roughly $5$ nm from the ion radially~\cite{Chen}.

In conclusion, the interaction between DNA and enzyme is realized through  water bridging the dynamics, and is more fully explained within the paradigm of gauge field theory: the interacting systems ``see'' each other by exchanging a mediating correlation dipole wave emerging from the em field pattern of the aromatic networks of DNA and enzyme molecules in their water environment. The experimental realization of DNA amplification in the PCR process described in Section 2 using signalized water (according to the protocol reported in Appendix A) may seem unconventional, though it is not altogether surprising given the complex molecular electrodynamics undergirding all of biochemistry. Thus, it is possible to envision that the chemical DNA template can be replaced by electromagnetic signals that register the collective dynamics of the DNA template's constituent charges and dipoles. We are aware that there are many open questions and problems to be solved in this arena. As such, detailed studies are warranted to further investigate the implications and theoretical underpinnings of these phenomena. Our results presented here on water bridging the dynamics of PCR, however, suggest the general applicability of the gauge theory paradigm of quantum fields to this area of inquiry.

\vspace{6pt}

\section*{Note added in proof}

After the publication of the present paper (submitted on 11 February 2017), we have read the paper "DNA's Chiral Spine of Hydration", by M. L. McDermott, et al., published on May 24, 2017 \cite{McDermott}
where the authors report their discovery of a chiral water superstructure surrounding DNA under ambient conditions. This is the first observation of a chiral spine of hydration templated by a biomolecule and has been obtained by use of chiral sum frequency generation (SFG) spectroscopy, a method analogous to circular dichroism measurements. In the specific case, water forms a robust chiral superstructure of the DNA helical structure.  The authors report that at room temperature and in 100 mM NaCl solution they "indeed observe that DNA imprints its chirality on the surrounding water molecules, generating a chiral SFG water response. This confirms the existence of a DNA minor groove spine of hydration at room temperature and further shows that the chiral structure of biomolecules can be imprinted on the surrounding solvation structure". The biological relevance of the discovery is evident.

Here we observe that such a discovery is fully consistent with the predictions presented in this paper, namely that polymerase chain reaction (PCR) processes rest on the mediating influence of such  hydration structures formed by the DNA template and Taq polymerase biomolecules . In fact, our conclusion is that, without organized water structures, no PCR processes can occur and no amplification of DNA can be obtained. The observation reported in \cite{McDermott} of the robustness of the DNA's chiral spine of hydration and the fact that "a change in the hydration state can lead to dramatic changes to the DNA structure" \cite{McDermott} also confirm that such a water superstructure actually constitutes a detailed mold or "electromagnetic image", as we have called it in our paper, of the DNA (and, additionally, other biomolecules) generating molecular electric dipole interactions. This electromagnetic image imprinted in the water dipole field is what the polymerase enzyme recognizes in the DNA's water environment. These studies open the way toward  novel  biomolecular  "imaging" technologies and new research disciplines, which have been incited by L. Montagnier's PCR experiments. The analysis in our paper, as well as in previous ones along similar research lines, is fully grounded in the quantum field theory description of the phenomenon and finds experimental support in this latest discovery of the existence of DNA's chiral spine of hydration \cite{addendum}.

\acknowledgments{Partial financial support from MIUR and INFN is gratefully acknowledged by AC, AP, PR and GV.  TJAC would like to acknowledge financial support from Nova Southeastern University's President's Faculty Research and Development Grant program (PFRDG 335426 - Craddock PI). PK would like to acknowledge financial support from the Whole Genome Science Foundation.}


\appendix

\newpage

\section{Protocol for DNA transduction in water}
In this Appendix we present briefly the main steps of the protocol for DNA transduction in water. Upon request, the protocol for DNA transduction in cells can be transmitted for research purposes only.

There are three different digital files in .wav format recorded at LM's  laboratories:\\
$\bullet$	 For {\it Borrelia} 16S DNA: Borr DNA16S.wav\\
$\bullet$	For HIV-1 LTR:\\
\indent 1.	LTRIn355S.wav\\
\indent 2.	LTRIn549AS.wav\\
$\bullet$	DW.wav, control of signalized water

The files have been shown in LM's laboratories to induce in water dilutions under proper conditions, the characteristic DNAs: 499 bp for {\it Borrelia} DNA; 194 bp for HIV-1 LTR~\cite{DNA1a,DNA1,DNA2a,DNA2}. Upon request, details on these files can be transmitted for research purposes only.

\subsection*{Preliminary step: how to use  the files received via Internet}

The computer (preferably Sony) must be equipped with an external  sound card (e.g. SoundBlaster, CREATIVELABS) comprising a digital-analog converter. The output  of the sound card is linked to the input of a commercial amplifier (Kool Sound SX-250)
having the following characteristics: passband from 10 Hz to 20 kHz, gain 1 to 20, input sensitivity 250 mV, output power RMS 140 W under 8 ohms.

The output of the amplifier is connected to a transducer solenoid which has the following characteristics: the bobbin has a length of 120 mm, an internal diameter of 25 mm, an external diameter of 28 mm .

\subsubsection*{1)	Preparation of water for transduction}

15 ml of DNAse/RNAse-free water (5-Prime, reference 2500010) in a 15 ml (polypropylene conical centrifuge tube, BD FalconTM REF 352096) are filtered first through a sterile 450 nm filter (Millex, Millipore, reference N. SLHV033RS) and then through 100 nm filter (Millex, Millipore, reference N. SLVV033RS), for {\it Borrelia} and through 20 nm filter (Whatman, Anotop 25, reference N. 6809-2002), for HIV-1 LTR.

\subsubsection*{2)	Signal transduction}

10 ml of this filtered water in a 15 ml (polypropylene conical centrifuge tube, BD FalconTM, reference 352096) are placed at the center of the solenoid, itself installed at room temperature on a non-conductive (non-metallic) working bench.

The modulated electric current produced by the amplifier is applied to the solenoid for 1 Hz at a  tension of 2 volts.
A current intensity of  1 A is applied to the coil, so that no significant heat is generated inside the cylinder.

\vspace{0.2cm}
\noindent 2.1) Playback (Upon request, details on accessing and using the downloaded files can be transmitted for research purposes only.)

The digital file is played for 1 hour.

\vspace{0.2cm}
\noindent 2.2) Transduction of the signalized water

By definition, the water which has received the specific recorded signal is called ``signalized water''.
The signalized water  (kept in the same tube) is shaken by strong vortex agitation for 15  seconds at room temperature.
1 ml is kept for control (NF, non-filtered), 3 mls of signalized water are filtered through a sterile 450 nm filter and then through a 100 nm filter. The filtrate is then diluted serially 1 in 10 (0.1 ml in 0.9 ml of DNAse/RNAse-free  water) in a DNA Lobind tube 1.5 ml (Eppendorf, reference 022431021) from $10^{-2}$ to $10^{-15}$ (D2 to D15). A strong vortex agitation is performed at each dilution step for 15 seconds.

\vspace{0.2cm}
\subsubsection*{ 3)	Reconstruction of the DNA by PCR}

5 $\mu$l of a signalized water dilution is added to 45 $\mu$l of the PCR mix, using PCRBIO HS \textit{Taq} DNA (PCRBIOSYSTEMS). This step is done for all the dilution samples.

\vspace{0.2cm}
\indent a)	Preparation of the mix for {\it Borrelia}
\vspace{0.2cm}

The PCR mixture (50 $\mu$l) contained 30.75 $\mu$l of DNAse/RNAse-free distilled water, 10 $\mu$l of 5 x PCRBIO reaction buffer, 2 $\mu$l of 10 $\mu$M solution of Forward primer and 2 $\mu$l of 10 $\mu$M solution of Reverse primer  [Borr16in-S $\&$ AS],
0.25 $\mu$l of 5 U/$\mu$l PCRBIO HS DNA Polymerase (PCRBIOSYSTEMS) and 5 $\mu$l of sample dilution as template. The PCR was performed with a mastercycler ep (Eppendorf). The PCR mixtures were incubated successively  at 61${}^{\circ}$C for 30 sec and 70${}^{\circ}$ C for 1 min followed by 40 PCR cycles of amplification (95${}^{\circ}$C for 15 sec; 61${}^{\circ}$C for 15 sec; 72${}^{\circ}$C for 45 sec). A final extension step was performed at 72${}^{\circ}$C for 10 min.

\vspace{0.3cm}
b)	Preparation of the mix for HIV-1 LTR
\vspace{0.2cm}

The PCR mixture (50 $\mu$l) contained 30.75 $\mu$l of DNAse/RNAse-Free distilled water, 10 $\mu$l of 5 x PCRBIO reaction buffer, 2 $\mu$l of 10 $\mu$M solution of Forward primer and 2 $\mu$l of 10 $\mu$M solution of Reverse primer [HIVLTRin-S \& AS],
0.25 $\mu$l of 5 U/$\mu$l PCRBIO HS \textit{Taq} DNA (PCRBIOSYSTEMS) and 5 $\mu$l of sample dilution as template. The PCR was performed with a mastercycler ep (Eppendorf). The PCR mixtures were incubated successively  at at 56${}^{\circ}$C for 30 sec and 70${}^{\circ}$C for 1 min followed by 40 PCR cycles of amplification (95${}^{\circ}$C for 15 sec; 56${}^{\circ}$C for 15 sec; 72${}^{\circ}$C for 45 sec). A final extension step was performed at 72${}^{\circ}$C for 10 min.

\vspace{0.4cm}

If DNA bands are faint with 40 cycles of amplification, it is advised to perform 30 more PCR cycles under the same conditions.

\vspace{0.4cm}
Electrophoresis of the PCR products in $1.3 \%$ agarose gel.

\vspace{0.2cm}
\subsubsection*{ 4) Sequencing}

The DNA bands are cut from the gel and purified using a Qiaquick gel extraction kit (Qiagen), and cloned in the pCR2.1-TOPO vector (Invitrogen). One Shot Mach1TM-T1R Chemically Competent {\it E. coli} are transformed with the ligated samples according to the supplier protocol (InVitrogen), spread onto X-gal and ampicillin-containing LB agar plates, and incubated at 37${}^{\circ}$C overnight. Single white colonies are picked and used to inoculate 4 ml of ampicillin-containing LB medium for plasmid production. The liquid bacterial minicultures are agitated (250 rpm) at 37${}^{\circ}$C overnight. Plasmid from each miniculture is purified with a QIAprep Spin Miniprep kit (Qiagen) and screened, after {\it Eco}RI digestion, for the presence of appropriate insert. The cloned DNA insert is then sequenced using a universal M13 primer. The sequence of the specific amplified DNA should be identical to the original DNA.
\vspace{0.5cm}

All the steps of the protocol described above need to be followed with much care and many precautions. Any change or omission in the main parameters, or  violation of any minor detail of the protocol, may result in failure of the experiment. In order to avoid em interferences in the measurement laboratory, sources of low-frequency EMS should be shielded, and cell phones should be turned off (with batteries removed) because some devices are regulated by low-frequency signals.

\newpage

\section{DNA and {\it Taq} dipole properties}

In this Appendix we report few dipolar properties of DNA and {\it Taq} molecules. Polarizabilities of DNA bases are given in Table 1.  The diagonal elements $\omega_{s,ii\,}$ of the angular frequency tensor for each base-pair electronic oscillator are given in Table 2 and are determined from polarizability data~\cite{watermediated,Kurian}.

\begin{table}[hp!]
\centering
\caption{Polarizabilities for DNA bases \cite{mcweeny1962perturbation, papadopoulos1988polarisability, basch1989electrical}, in units of 1 au = $1.64878 \times 10^{-41} \text{C}^2 \text{m}^2 \text{J}^{-1}$. Reproduced from Ref.~\cite{watermediated}.}
\label{DNAalpha}
\begin{tabular}{ c  c  c  c}
\hline
\hline
DNA base & $\alpha_{xx}$ & $\alpha_{yy}$ & $\alpha_{zz}$  \\ \hline
Adenine (A) & 102.5  & 114.0 & 49.6 \\
Cytosine (C) & 78.8  & 107.1 & 44.2 \\
Guanine (G)  & 108.7 & 124.8 & 51.2\\
Thymine (T) & 80.7    & 101.7  & 45.9\\
\hline \hline
\end{tabular}
\end{table}

\begin{table}[hp!]
\centering
\caption{DNA base pair (bp) electronic angular frequencies, calculated from polarizability data, in units of $10^{15}$ radians per second. Reproduced from Ref.~\cite{Kurian}.}
\label{DNAbpomega}
\begin{tabular}{ c  c  c  c }
\hline
\hline
bp & $\omega_{xx}$ & $\omega_{yy}$ & $\omega_{zz}$ \\ \hline
A:T & 3.062 & 2.822 & 4.242\\
C:G & 3.027 & 2.722 & 4.244\\ \hline \hline
\end{tabular}
\end{table}

\vspace{0.2cm}
Aromatic amino acid polarizabilities and electronic angular frequencies are given in Tables \ref{Trpalpha} and \ref{Trpomega}, respectively.

\begin{table}[hp!]
\centering
\caption{Aromatic amino acid polarizabilities \cite{indole, phenol, benzene1, benzene2, benzene3}, in units of 1 au = $1.64878 \times 10^{-41} \text{C}^2 \text{m}^2 \text{J}^{-1}$, with $\overline{\alpha} = \sqrt{\alpha_{xx}^2 + \alpha_{yy}^2 + \alpha_{zz}^2}$. Reproduced from Ref.~\cite{watermediated}.}
\label{Trpalpha}
\begin{tabular}{ ccccc }
\hline
\hline
Amino Acid & $\overline{\alpha}$ & $\alpha_{xx}$ & $\alpha_{yy}$ & $\alpha_{zz}$ \\ \hline
Trp & 153.4 & 119.5 & 91.6 & 29.4 \\
Tyr & 129.3  & 89.5 & 43.0 & 82.9 \\
Phe &118.1 & 79.0 & 79.0 & 38.6 \\
\hline \hline
\end{tabular}
\end{table}

\vspace{0.2cm}

By following Ref. \cite{Kurian}, the normal-mode lowering operator is introduced as
\begin{equation}  \nonumber
a_{s, j}=\sqrt{\frac{m\Omega_{s,j}}{2\hbar}}(\v{r}^\prime_s)_j+ \frac{i}{\sqrt{2m \hbar \Omega_{s,j}}}(\v{p}^\prime_s)_j,
\end{equation}
where $s=0,1,\dots,N-1$, $j=xy, z$ and
\begin{equation} \nonumber
(\v{r}^\prime_s)_j = \sum \limits_{n=0}^{N-1} (\v{r}_n)_j \, \exp\left(-\frac{2\pi i ns}{N}\right), \qquad \quad
(\v{p}^\prime_s)_j = \sum \limits_{n=0}^{N-1} (\v{p}_n)_j \, \exp \left(-\frac{2\pi i ns}{N}\right).
\end{equation}

\begin{table}[hp!]
\centering
\caption{Aromatic amino acid electronic angular frequencies, calculated from polarizability data, in units of $10^{15}$ radians per second. Reproduced from Ref.~\cite{watermediated}.}
\label{Trpomega}
\begin{tabular}{ c c c c c }
\hline
\hline
Amino Acid & $\overline{\omega}$ & $\omega_{xx}$ & $\omega_{yy}$ & $\omega_{zz}$   \\ \hline
Trp & 3.338 & 3.782 & 4.320 & 7.622 \\
Tyr & 3.635 & 4.370 & 6.305 & 4.541\\
Phe & 3.803 & 4.653 & 4.653 & 6.654 \\
 \hline \hline
\end{tabular}
\end{table}

\vspace{0.2cm}

The collective eigenmode frequencies for the dipole oscillations can be obtained for a range of biopolymers, including nucleic acids and other proteins~\cite{watermediated,Kurian}. At an elementary level, however, the longitudinal mode frequencies $\Omega_{s,z}$ in the four-bp case take on the following simple form when DNA sequence homogeneity is assumed, so that by making the replacements 
$\omega_{s,zz}=\omega$, and $\gamma_{s,s+1}^z = \gamma=-Q^2/(2\pi\epsilon_0 d^3)$, we obtain:
\begin{equation}  \nonumber 
\Omega_{0,z}^{\,2} = \omega^2-\varphi \frac{\gamma}{m};  \qquad ~
\Omega_{1,z}^{\,2} = \omega ^2- (\varphi-1)\frac{\gamma }{m}; \qquad ~
 \Omega_{2,z}^{\,2} = \omega ^2+ (\varphi-1)\frac{\gamma }{m}; \qquad ~
 \Omega_{3,z}^{\,2} = \omega ^2+ \varphi \frac{\gamma}{m},
\end{equation}
where $\varphi = \left(1+\sqrt{5}\right)/2$ is the golden ratio. See Ref.~\cite{Kurian} for further details.

\newpage

\section{The water dipole field}

By using the notation of Refs.~\cite{watermediated} and \cite{PRA2006},  let $\phi ({\bf x}, t)$ be the complex dipole wave field collectively describing the system of $N$ water molecules in the unit volume $V$, so that integration over the sphere of radius $\bf r$  gives:
\begin{equation} \nonumber 
\int d\Omega |\phi ({\bf x},t)|^{2} = N~, \end{equation}
where $d\Omega = \sin \theta d \theta d \phi$ is the element of solid angle and $(r, \theta, \phi)$ denote polar coordinates.
In Section 3 we have introduced the rescaled field  $\chi ({\bf x},t) = \frac{1}{\sqrt{N}} \phi ({\bf x},t)$.  The normalization over the unit sphere is
\begin{equation} \nonumber 
 \int d\Omega |\chi ({\bf x},t)|^{2} = 1~.
\end{equation}
The average value of the moment of inertia $I$ of water considered in the text has been chosen considering that it varies within a factor of three depending on the axis around which it is calculated: $1.0220 \times 10^{-40}$ g$\cdot$cm$^2$ for the axis in the plane of the water molecule with the origin on the oxygen and orthogonal to the H-O-H angle, i.e., parallel to the longest dimension of the molecule; $2.9376 \times 10^{-40}$ g$\cdot$cm$^2$ for the axis orthogonal to the plane of the water molecule with the origin on the oxygen; and $1.9187 \times 10^{-40}$ g$\cdot$cm$^2$ for the axis in the plane of the water molecule with the origin on the oxygen and bisecting the H-O-H angle (see Ref.~\cite{Eisenberg}).

As observed in the text, due to the dipole rotational invariance the amplitude of $\alpha_{1,m}(t)$ in Eq.(\ref{9}) does not depend on $m$, and  the normalization condition over the unit sphere then becomes
\be \nonumber 
|\al_{0,0}(t)|^{2} + \sum_{m} |\al_{1,m}(t)|^{2} =
|a_{0}(t)|^{2} + 3 |a_{1}(t)|^{2}   = 1 , \ee
at any time $t$. By putting  $Q \equiv |a_{0}(t)|^{2} + 3 ~|a_{1}(t)|^{2}$, we find that ${\pa}Q/{\pa t} = 0$. This expresses the conservation of the total number $N$ of molecules. Moreover, at each time $t$ the rate of change of the population in each of the levels $(1,m)$ for $m = 0, \pm 1$, equally contributes, in the average,  to the rate of change in the population of the level $(0,0)$, namely that
\be \nonumber 
\frac{\pa}{\pa t}{|a_{1}(t)|^{2}} = -
\frac{1}{3}~\frac{\pa}{\pa t}{|a_{0}(t)|^{2}}. ~\ee
One finds~\cite{PRL1988,PRA2006,Blasone2011} that another constant of motion is given by
\be \nonumber 
|u(t)|^{2} + 2~|a_{1}(t)|^{2} =
\frac{2}{3}\sin^{2}\theta_{0}. \ee
Since $|u(t)|^{2} \geq 0$, we have that
$|a_{1}(t)|^{2} \leq \frac{1}{3}\sin^{2}\theta_{0}$ and therefore 
$|a_{0}(t)|^{2} \geq \cos^{2}\theta_{0}$.

\newpage

\bibliographystyle{mdpi}

\renewcommand\bibname{References}

\end{document}